\documentclass[journal=ancac3,manuscript=article]{achemso}

\usepackage[version=3]{mhchem} 
\usepackage{amsmath,amssymb,mathrsfs,accents}
\usepackage{color}
\usepackage[scr=rsfso]{mathalpha}
\usepackage{booktabs}                   
\usepackage{multirow}                   
\usepackage{diagbox}                    
\usepackage{bm}
\usepackage{physics}
\usepackage[T1]{fontenc}
\usepackage[utf8x]{inputenc}
\usepackage{float}
\usepackage{graphicx}

\author{Xiao-ke Zhu}
\affiliation{Department of Applied Physics and Science Education, Eindhoven University of Technology, 5600MB Eindhoven, The Netherlands}
\alsoaffiliation{Centre for Optical and Electromagnetic Research, National Engineering Research Center for Optical Instruments, Zhejiang University, Hangzhou 310058, China}
\author{Yu-Chen Wei}
\email{y.c.wei@tue.nl }
\affiliation{Department of Applied Physics and Science Education, Eindhoven University of Technology, 5600MB Eindhoven, The Netherlands}
\author{Jose L. Pura}
\affiliation{Instituto de Estructura de la Materia (IEM-CSIC), Consejo Superior de Investigaciones Científicas, Serrano 121, 28006 Madrid, Spain.}
\alsoaffiliation{GdS-Optronlab, Física de la Materia Condensada, Universidad de Valladolid, Paseo de Belén 19, 47011 Valladolid, Spain.}
\author{Matthijs Berghuis}
\affiliation{Department of Applied Physics and Science Education, Eindhoven University of Technology, 5600MB Eindhoven, The Netherlands}
\author{Minpeng Liang}
\affiliation{Department of Applied Physics and Science Education, Eindhoven University of Technology, 5600MB Eindhoven, The Netherlands}
\author{Beatriz Castillo López de Larrinzar}
\affiliation{Instituto de Micro y Nanotecnología IMN-CNM, CSIC, CEI UAM+CSIC, Isaac Newton 8, E-28760 Tres Cantos, Madrid, Spain.}
\alsoaffiliation{Instituto de Estructura de la Materia (IEM-CSIC), Consejo Superior de Investigaciones Científicas, Serrano 121, 28006 Madrid, Spain.}
\author{Shunsuke Murai}
\affiliation{Department of Electronics and Physics, Graduate School of Engineering, Osaka Metropolitan University, Osaka, 599-8531, Japan}
\author{Antonio García-Martín}
\affiliation{Instituto de Micro y Nanotecnología IMN-CNM, CSIC, CEI UAM+CSIC, Isaac Newton 8, E-28760 Tres Cantos, Madrid, Spain.}
\author{José A. Sánchez-Gil}
\affiliation{Instituto de Estructura de la Materia (IEM-CSIC), Consejo Superior de Investigaciones Científicas, Serrano 121, 28006 Madrid, Spain.}
\author{Sailing He}
\affiliation{Centre for Optical and Electromagnetic Research, National Engineering Research Center for Optical Instruments, Zhejiang University, Hangzhou 310058, China}
\author{Jaime G\'omez Rivas}
\email{j.gomez.rivas@tue.nl}
\affiliation{Department of Applied Physics and Science Education, Eindhoven University of Technology, 5600MB Eindhoven, The Netherlands}

\title[An \textsf{achemso} demo]
{Robust Circularly Polarized Luminescence via Quasi-Bound States in the Continuum in Intrinsic Chiral Silicon Metasurfaces}

\begin{document}
\begin{tocentry}
\centering
    \includegraphics[scale=0.48]{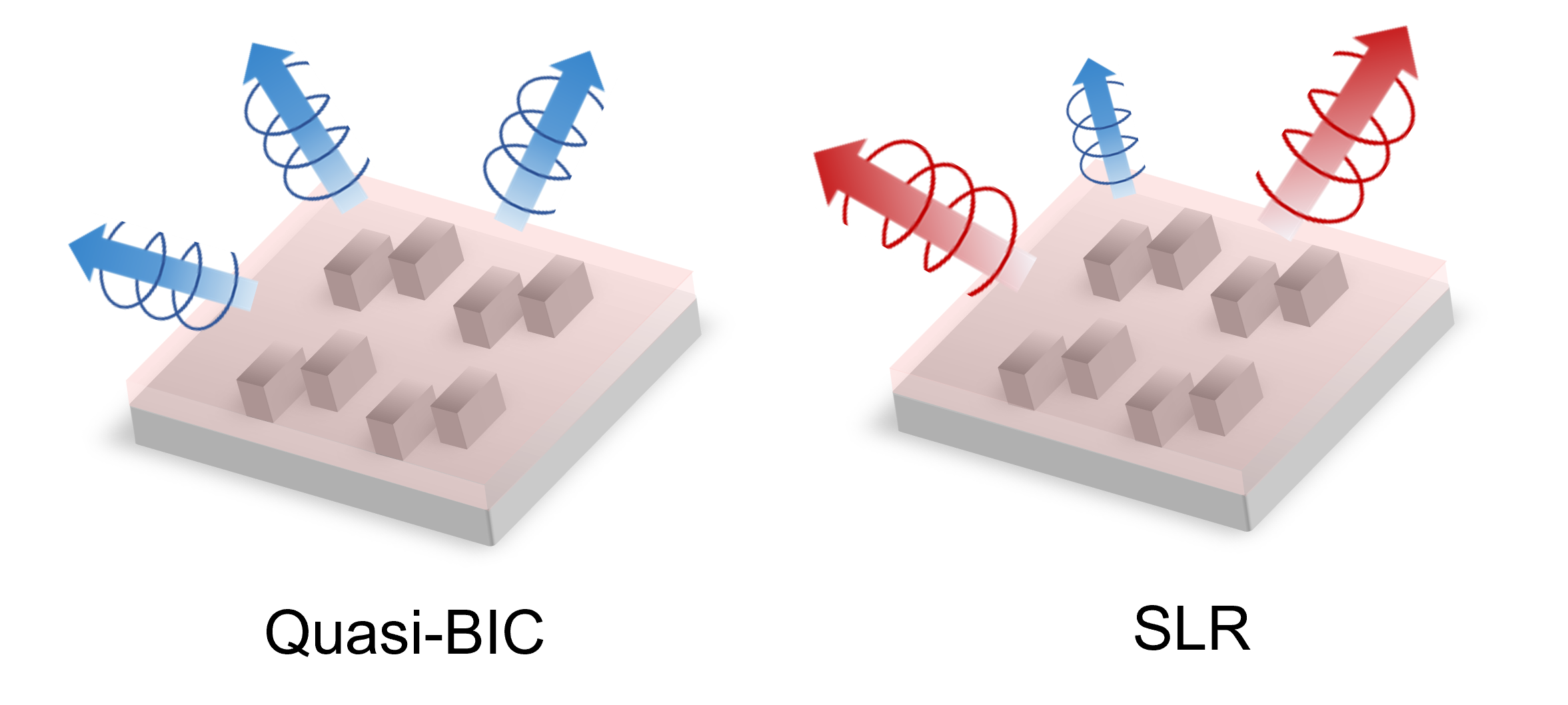}

\end{tocentry}


\begin{abstract}
  We demonstrate a circularly polarized photoluminescence emission, with dissymmetry factors $g_\mathrm{PL}$ over 0.1, from achiral organic dye molecules by leveraging quasi-bound states in the continuum (quasi-BICs) and surface lattice resonances (SLRs) in intrinsic silicon chiral metasurfaces. We find that the $g_\mathrm{PL}$ associated with the quasi-BIC mode remains robust against variations in emission angle and dye thickness owing to its strong lateral field confinement. In contrast, the $g_\mathrm{PL}$ of the SLR mode exhibits sign inversion depending on the emission energy and dye layer thickness. The experimental results are supported by mode decomposition analysis, helicity density analysis, and near-field spatial distribution of the electric field. These findings illustrate the relevance of the emitter's layer thickness in optimizing the emission of circularly polarized light. They also elaborate on the robustness of chiral quasi-BICs, offering insights into chiral light-matter interactions and advancing the design of circularly polarized light-emitting devices.
\end{abstract}

\section{Introduction}
Objects that cannot be superimposed on their mirror images by translations and rotations exhibit intrinsic chirality, which can be observed not only in matter but also in the polarization states of light~\cite{longhi2016circularly,lininger2023chirality}. Circularly polarized light, a manifestation of chirality in light,  offers additional degrees of freedom for encoding information compared to linearly polarized light. This advantage makes chiral luminophores exhibiting circularly polarized luminescence (CPL) particularly interesting for advanced technologies, including 3D displays and augmented reality, optical data storage, optical communication, metrology, imaging, microscopy, and medical diagnostics~\cite{lodahl2017chiral,mohammadi2018nanophotonic,ayuso2019synthetic,long2020chiral,seo2021circularly,kim2023chiral,deng2024advances,han2024chiral,wang2024automatically}. The degree of circular polarization in light can be quantified by the photoluminescence dissymmetry factor, $g_\mathrm{PL}$, defined as $g_\mathrm{PL}=2(I_\mathrm{LCP}-I_\mathrm{RCP})/(I_\mathrm{LCP}+I_\mathrm{RCP})$, where $I_\mathrm{LCP}$ and $I_\mathrm{RCP}$ represent the intensities of left-handed and right-handed CPL, respectively~\cite{wakabayashi2014anisotropic,vanorman2025chiral}.  Traditionally, sources of CPL are fabricated using bulky optical elements, such as polarizers and quarter-wave plates \cite{liang2016integratable,okazaki2023luminescence}, which are unsuitable for micro or nano-scale device platforms. Direct generation of chiral emission from materials offers a pathway toward compact device development. Materials such as chiral perovskites, small chiral organic dyes, and metal complexes have shown promise for generating circularly polarized emission \cite{deng2021circularly,crassous2023materials,greenfield2021pathways}. However, these materials often face challenges including low $g_\mathrm{PL}$ or low internal/external quantum efficiencies. Nowadays, achieving high $g_\mathrm{PL}$ and quantum efficiencies at room temperature remains challenging~\cite{deng2021circularly,crassous2023materials,greenfield2021pathways}. 

To address this challenge, various types of chiral nanostructures can be applied to enhance $g_\mathrm{PL}$, including plasmonic, excitonic, self-assembled, and photonic systems~\cite{hu2022plasmonic,lannebere2024chiral,hentschel2017chiral,lv2022self,anantharaman2021exciton,wu2023plasmon}. Among these, metasurfaces have emerged as a versatile platform for chiro-optical studies, offering unprecedented design flexibility compared to natural materials~\cite{deng2024advances}. The highly enhanced chiral light–matter interactions in metasurfaces are crucial for applications such as enantiomer selection, chiral molecular sensing, and chiral quantum optics~\cite{farshchi2011optical,li2024metasurface,deng2024advances,sha2024chirality}. Periodic arrays of nanoparticles in metasurfaces can induce a range of optical phenomena through the radiative coupling of local dipoles and multipoles in each particle. For example, surface lattice resonances (SLRs) arise from the hybridization between localized resonances and Rayleigh anomalies \cite{Vecchi2009,wang2019manipulating,kravets2018plasmonic}, which have been utilized to achieve significant chiral photoluminescence in quantum dots, perovskites, and molecular dyes~\cite{Cotrufo2016,wang2019observation,mendoza2023nanoimprinted,liu2023controlling,fiuza2024inducing,mendoza2024single,hong2025high,qi2025chiral}. Additionally, if the particle arrays restrict the available radiation channels, these modes remain localized to the structure even though they coexist with the continuum of radiative modes, giving rise to bound states in the continuum (BICs) with a theoretically infinite temporal field confinement \cite{marinica2008bound,hsu2016bound}. In addition, by introducing in-plane asymmetry, BICs turn into quasi-BICs with finite but still large field confinements. These optical modes enable significant chiral photoluminescence (CPL) as well as applications in chiral sensing~\cite{gorkunov2020metasurfaces,overvig2021chiral,zhang2022chiral,shi2022planar,sun2023enhancing,lim2023maximally,liang2024tailoring,khaliq2023recent,kim2025ultranarrowband}. 

In this manuscript, we investigate the CPL from an achiral organic dye enabled by SLRs and quasi-BICs in intrinsic chiral silicon metasurfaces. Leveraging these modes, we achieve $g_\mathrm{PL}$ exceeding 0.1. The quasi-BIC mode exhibits robust CPL with a consistent $g_\mathrm{PL}$ of 0.1, remaining stable across variations in dye thickness and emission angle. In contrast, the SLR mode demonstrates a higher $g_\mathrm{PL}$ of 0.17. Notably, the sign of $g_\mathrm{PL}$ within the same SLR mode can be inverted by altering the dye thickness or emission angle. These behaviors are elucidated through analysis of the local electric field distribution, the helicity density, and a modal decomposition analysis. Our findings highlight the comparative robustness of CPL between SLR and quasi-BIC modes in intrinsic chiral metasurfaces, paving ways for the development of advanced chiral luminescent devices with robust emission.

\begin{figure}
    \centering
    \includegraphics[width=0.8\linewidth]{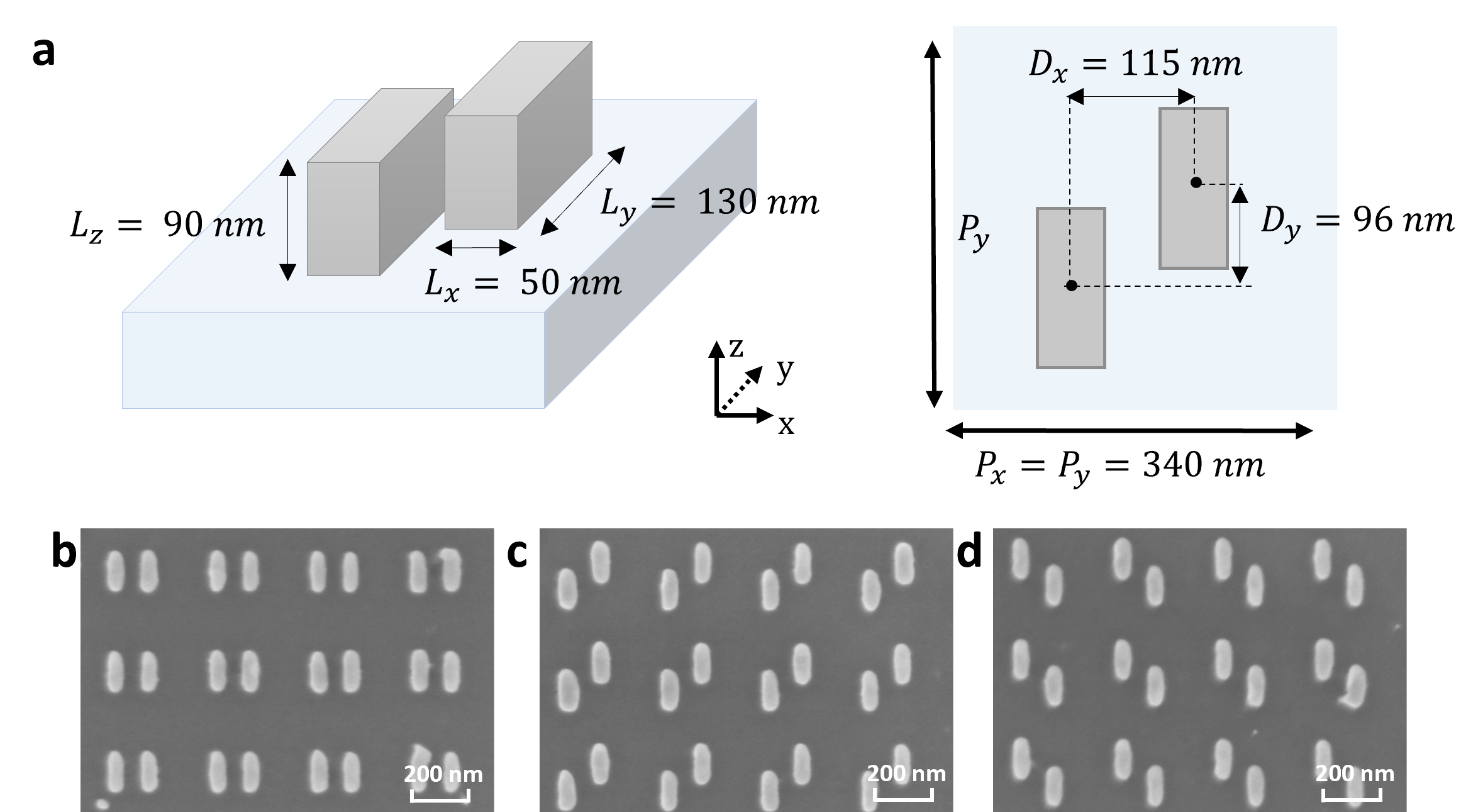}
    \caption{Chiral metasurfaces formed by arrays of silicon nanorod dimers. (a) Scheme of one unit cell of the chiral metasurface composed of two displaced Si nanorods. The structural parameters are indicated. The dotted line of the y-axis indicates the direction into the plane. (b-d) SEM images of three metasurfaces with unit cells formed by (b) equal non-vertically displaced Si nanorods (achiral), (c) positively displaced Si nanorods in the y-direction, and (d) negatively displaced Si nanorods in the y-direction.}
    \label{Fig1}
\end{figure}

\textit{Sample description and photoluminescent enhancements.} The metasurfaces consist of a matrix of polycrystalline silicon nanorod dimers with height L$_z= 90$ nm, width L$_x= 50$ nm, and length L$_y=130$ nm on a glass substrate, placed in a square lattice with a unit cell of $340\times340$ nm$^2$ (see Figure~\ref{Fig1}a for a schematic illustration). For the configuration with the maximum symmetry (Figure~\ref{Fig1}b), the distance between the two rods along the x-axis  (D$_x= 115$ nm) is large enough to analyze the two rods within the dipolar approximation. To manipulate the chiro-optical activity, we study the intrinsic dissymmetry arising from the displacements of one nanorod along the y-axis (D$_y=\pm96$ nm) (Figures~\ref{Fig1}c and \ref{Fig1}d). Note that the unit cell with the opposite sign of D$_y$ is an enantiomorph. As a result of symmetry restrictions, we expect the chiro-optical responses from these enantiomorphs to be inverse of each other. The sample area for each sample is 2 $\times$ 2 mm$^2$. The details of the sample fabrication can be found in the Methods section. Figures~\ref{Fig1}b-d show the scanning electron microscopy (SEM) images of the samples, which illustrate the designed structural dimensions.

\begin{figure}[H]
    \centering
    \includegraphics[width=0.9\linewidth]{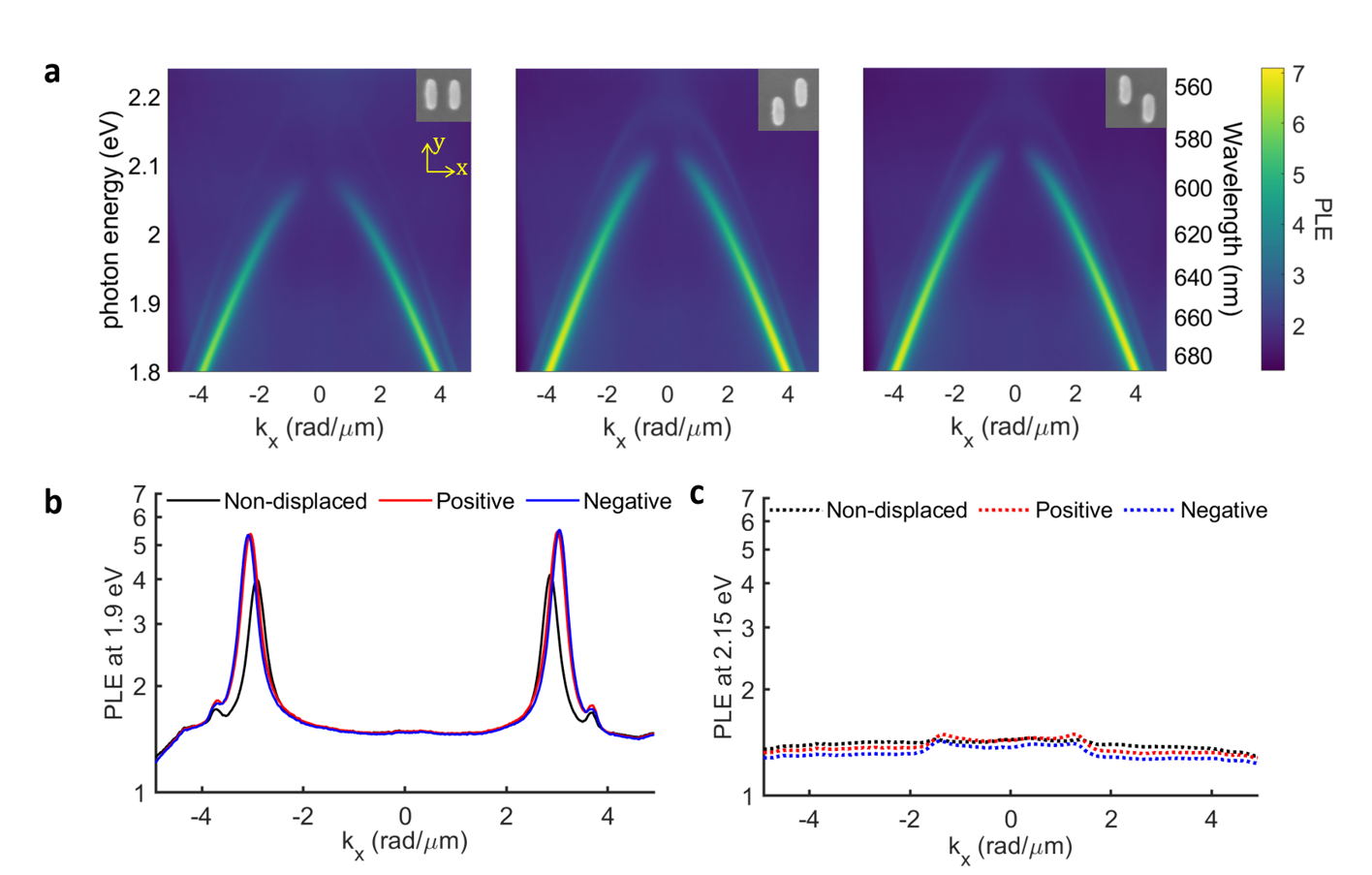}
    \caption{Photoluminescence (PL) analysis for three metasurfaces of nanorod dimers with a spin-coated dye-doped polymer film on top with a thickness of 220 nm.  (a) Photoluminescence enhancement (PLE) for metasurfaces with an unit cell formed by two equal non-vertically displaced Si nanorods (left panel), two positively displaced Si nanorods in the y-direction (central panel), and two negatively displaced Si nanorods (right panel) in the y-direction, measured with a Fourier microscope as the function of the in-plane wave vector $k_x$. The insets show the SEM images of the unit cell for each structure. PLE as a function of $k_x$ at (b) 1.9 eV and (c) 2.15 eV represented with solid and dotted curves, respectively.}
    \label{Fig2}
\end{figure}

After fabrication of the metasurfaces, we spin-coated on top a 220 nm thick layer of 5 wt\% by weight perylene dye ([N, N'-bis (2,6-diisopropylphenyl) -1,7- and -1,6-bis (2,6-diisopropylphenoxy) perylene -3,4: 9,10-tetracarboximide]) in poly(methyl methacrylate) (PMMA). The bare perylene dye shows two exciton peaks in both the extinction and the photoluminescence (PL) spectra, corresponding to the electronic transition and its first vibronic replica, respectively (see Figure~S1 in the supplemental material). We use a continuous-wave laser with a wavelength of  532 nm to excite the three samples and measure their angle-dependent PL enhancement (PLE) with a Fourier microscope. The PLE is analyzed as a function of the emission wavelength and the in-plane wavevector (parallel to the surface) along the x-axis ($k_x$). The Fourier microscope used for these measurements is schematically shown in Figure~S2 of the supplemental material. The PLE is obtained from the PL maps of the dye on metasurfaces divided by the PL maps measured of a bare organic layer. To avoid polarized excitation from the laser source, we add an optical diffuser in front of the sample to achieve unpolarized excitation. In Figure~\ref{Fig2}, the PLE dispersion measurements are shown for the parallel wave vector along the short axis of the nanorods (x-axis). The dispersion along the x-axis shows two distinct modes. The mode with larger enhancements (PLE > 5) corresponds to a BIC (Figure~\ref{Fig2}a). 
The PLE vanishes in the direction normal to the surface ($k_x=0$) due to the inversion symmetry of the arrays along this direction. This inversion symmetry is the origin of the symmetry protection, leading to the suppression of far-field emission~\cite{tikhodeev2002quasiguided,lee2012observation,van2021unveiling}. For wave vectors other than 0, the inversion symmetry is broken and radiation leakage is possible, giving rise to the quasi-BIC modes. By contrast, the mode with small enhancements (PLE $\approx 1.5$) and radiation leakage at normal incidence in these dispersion measurements corresponds to SLR. The stronger PLE from the quasi-BIC is associated with its higher Q-factor and field confinement to the surface \cite{watanabe2024low,ter2023direct}. Note that exhibiting both SLR and quasi-BIC on the same array offers an effective platform for comparing their photonic characteristics. The PLE measurements with in-plane wave vectors along the y-axis are shown in Figure~S3 in the supplementary information. There is only one SLR mode visible with significant PLE (over a factor 8), and one low-dispersion band at 2.1 eV with a weak PLE $\approx 1$. 



\begin{figure}[H]
    \centering
    \includegraphics[width=0.9\linewidth]{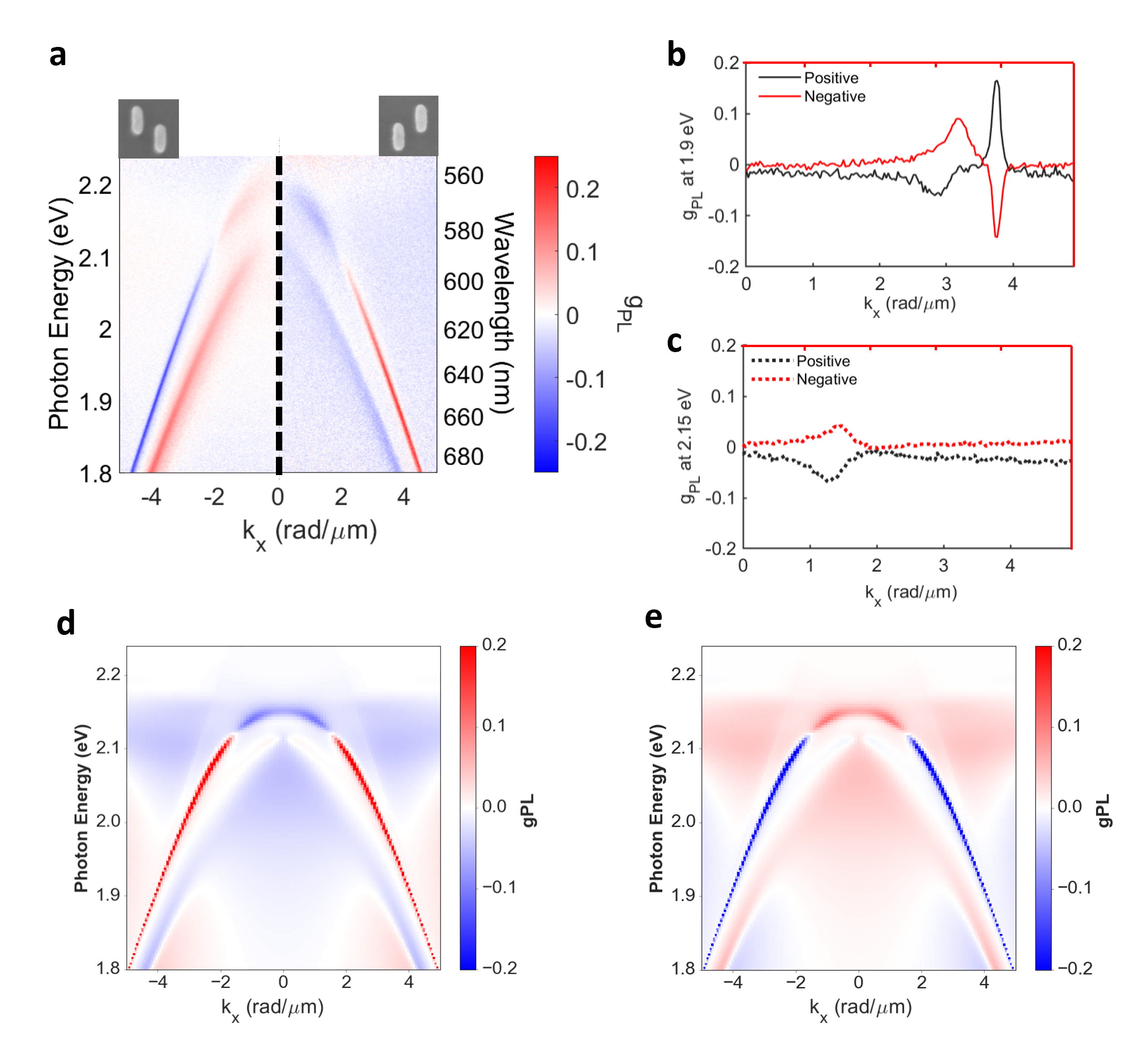}
    \caption{PL dissymmetry factor of the chiral metasurfaces with a 220 nm thick layer of dye-doped polymer on top. $g_\mathrm{PL}$ maps as a function of photon energy (wavelength) and $k_x$ for the array of positively displaced Si nanorods in the y-direction (right panel), and the negatively displaced Si nanorods (left panel). $g_\mathrm{PL}$ as the function of $k_x$ at (b) 1.9 eV and (c) 2.15 eV. Calculated PL dissymmetry maps of (d) the positively displaced rods and (e) the negatively displaced rods radiating along the x axis with a 220 nm dye-doped polymer film.
}
    \label{Fig3}
\end{figure}

 \textit{Chiral emission measurements.} To characterize the degree of circular polarization, we measure the circularly polarized PLE maps by placing a quarter-wave plate and a linear polarizer in front of the detector (Figure~S2 in the supplementary information). Using the PLE maps of right-handed circular polarization (RCP) and left-handed circular polarization (LCP), we determine the $g_\mathrm{PL}$ maps, as illustrated in Figure~\ref{Fig3}. For the non-displaced sample, the $g_\mathrm{PL}$ value is zero due to its achiral structure (Figure~S4 in the supplementary information). In addition, the mirror symmetry between positively and negatively displaced rods in the unit cell results in $g_\mathrm{PL}$ dispersion maps with opposite signs and similar magnitudes (note that the left and right panels of Figure~\ref{Fig3}a  correspond to the dispersion of $g_\mathrm{PL}$ for negatively and positively displaced rods, respectively). For the emission along the x-axis, the SLR mode produces a significant $|g_\mathrm{PL}|$ of approximately 0.17, while the quasi-BIC mode exhibits a relatively smaller $|g_\mathrm{PL}|$ value of approximately 0.10 at its peak. For the emission along the y-axis (Figure~S5 in the supplementary information), the SLR peak exhibits similar $|g_\mathrm{PL}|\approx 0.2$, while the low-dispersion mode shows a maximum $|g_\mathrm{PL}|\approx 0.1$. Interestingly, in both emission directions, the $g_\mathrm{PL}$ measured at 1.9 eV has the opposite sign of the $g_\mathrm{PL}$ measured at 2.15 eV (Figure~\ref{Fig3}b-\ref{Fig3}c). It is worth noting that a similar array structure made of metallic materials has exhibited higher $g_\mathrm{PL}$ under high excitation powers due to nonlinear (lasing) emission, in contrast to the present study, which is conducted in the linear response regime.

\textit{
Discussion.} To elaborate on the far-field emission enhancement, we performed numerical calculations based on COMSOL. The PL dissymmetry factor is obtained from the scattering problem illuminating with circularly polarized light and invoking Kirchhoff's law, as explained in the Methods. The resulting $g_\mathrm{PL}$ bands for different frequencies and angles of incidence are shown in Figure~\ref{Fig3}d-\ref{Fig3}e. 
The agreement with the experimental measurements is notable, showing only a slight frequency shift. Nonetheless, the measured  $g_\mathrm{PL}$ for the quasi-BIC band is somewhat larger (especially for the negatively displaced nanorods) than that numerically calculated. We associate the discrepancy to sample imperfections that may induce a larger PL dissymmetry for the BIC band.

\begin{figure}
    \centering
    \includegraphics[width=0.9\linewidth]{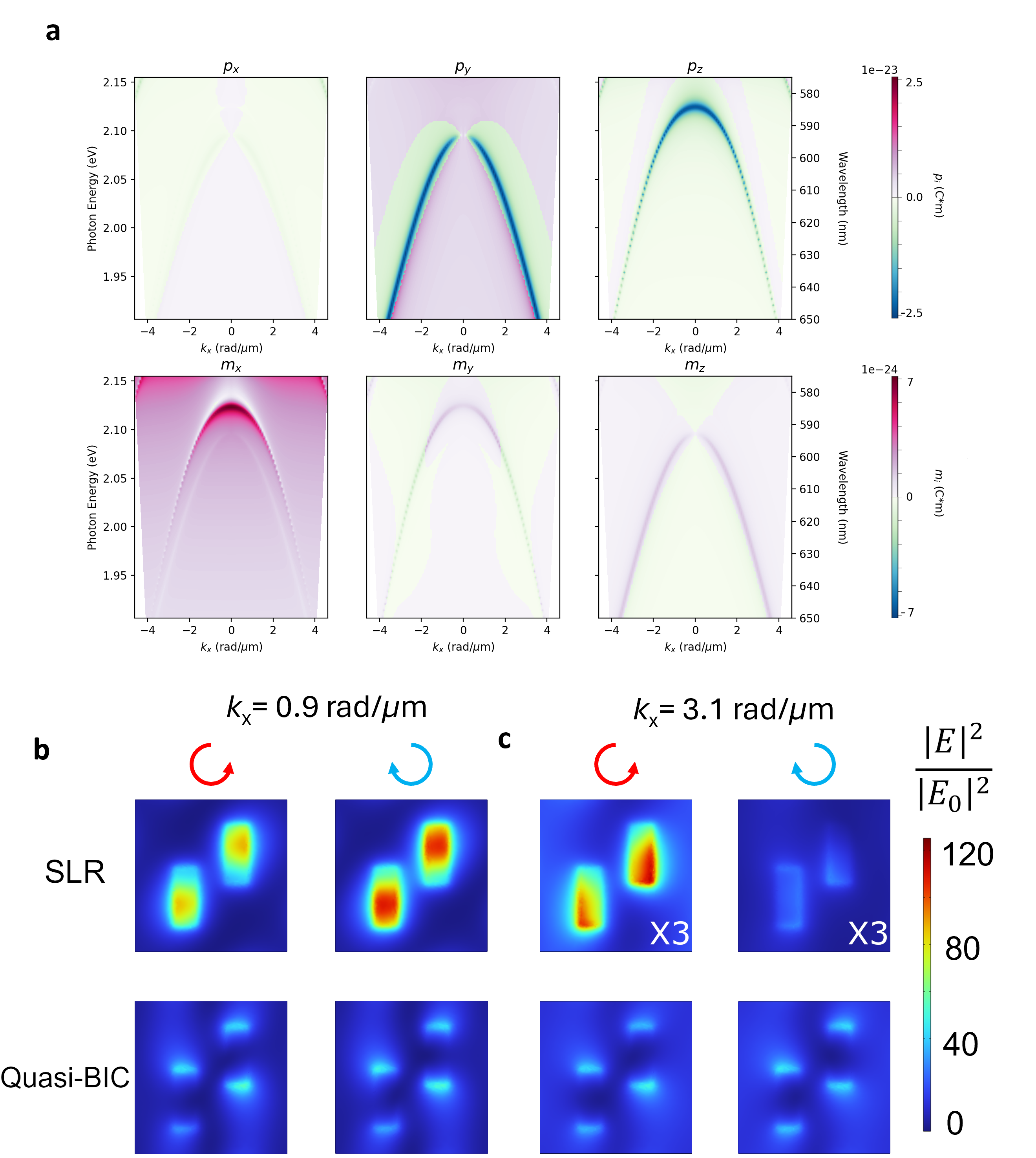}
    \caption{(a) Angle-dispersive mode contribution maps simulated along $k_x$ axis of the array of positively displaced Si nanorods with the dye film thickness of 220 nm. Upper panels: electric dipole components along the $x,y$ and $z$ directions; lower panels: magnetic dipole components along the $x,y$ and $z$ directions. The magenta color scale indicates the contribution of symmetric dipoles per unit cell, while the blue/green color scale indicates the contribution of antisymmetric dipoles. (b) and (c) Electric near-field distribution of the SLR mode and the quasi-BIC mode on the $xy$ plane at a height of 10 nm over the top of the rods with a 220 nm dye-doped polymer film with an illumination at (b) $k_x$ = 0.9$~\mathrm{rad}/\mu\mathrm{m}$ (2.11 eV) and (c) 3.1$~\mathrm{rad}/\mu\mathrm{m}$ (1.98 eV).}
    \label{Fig4}
\end{figure}

 In order to shed light on the chirality of the modes, a multipolar decomposition of the induced near-fields by the metasurfaces is also carried out through COMSOL (see  Methods). We only consider the contribution of the electric ($\textbf{p}$) and the magnetic ($\textbf{m}$) dipole moments, since previous studies indicate that the quadrupolar response extends beyond the spectral region we focus on~\cite{pura2024superchiral}. For simplicity, we only analyzed the results of the positively displaced rods since the physical interpretation in the negatively displaced rods must be symmetric. We analyzed the electric and magnetic dipoles of each rod: $(\mathbf{p_1}, \mathbf{m_1})$ and $(\mathbf{p_2},\mathbf{m_2})$. 
However, according to the $C_2$ (inversion) symmetry of the system, it is sufficient to analyze the magnitude of the dipoles in one rod, since $\rm{p}=|\mathbf{p_1}|=|\mathbf{p_2}|$ and $\mathbf{m}=|\mathbf{m_1}|=|\mathbf{m_2}|$, and the symmetric/antisymmetric character of the pair, as the only two possibilities are $\mathbf{p_1} = \mathbf{p_2}$ and $\mathbf{p_1} = -\mathbf{p_2}$ for the electric dipoles, and $\mathbf{m_1} = \mathbf{m_2}$ and $\mathbf{m_1} = -\mathbf{m_2}$ for the magnetic dipoles. Both p and m are presented in units of Coulomb times meter $(C\cdot m)$ for direct comparison. For this m is normalized by the factor $n/c$, where $n$ is the refractive index of the medium surrounding the metasurface and $c$ the speed of light in vacuum. The results for the multipolar contributions to reflection upon illuminating with a wave vector  along the x-axis are presented in Figure~\ref{Fig4}a for the case of the positively displaced rods.  We exploit the dipolar symmetries mentioned above, plotting the magnitude of the dipolar contributions along the cartesian directions for a single nanorod, encoding in the color the symmetric/antisymmetric character in the dimer: the magenta color scale indicates the contribution of symmetric dipoles per unit cell, while the blue/green color scale indicates the contribution of antisymmetric dipoles. 
Figure~\ref{Fig4}a shows that the quasi-BIC mode is associated with a significant in-plane antisymmetric $\rm{p}_y$ mode and a weak out-of-plane symmetric $\rm{m}_z$, both non-radiative at $k=0$. The resulting $\pi$-rotation symmetry confirms the BIC character at the $\Gamma$-point with diverging Q-factor, which supports its high PLE along the quasi-BIC band.  The direction of $\textbf{p}$ is verified by the experimental extinction maps, showing that the quasi-BIC mode is associated with the excitation polarization along the y-axis (Figure~S6 in the supplementary information). The antisymmetric $\rm{p}_y$  character of the dimer in turn supports the emergence of positive (respectively, negative) $g_\mathrm{PL}$ for negatively (respectively, positively) displaced nanorod in Figure~\ref{Fig3}a.
On the other hand, the SLR mode arises from the out-of-plane antisymmetric $\rm{p}_z$ together with a strong in-plane symmetric $\rm{m}_x$ contribution. This mode cannot become a BIC at the  $\Gamma$-point due to the lack of $\pi$-rotation symmetry within the unit cell and the in-plane character of the $m_x$ contribution.
Note that the significant $m_x$ in the SLR mode explains its high $g_\mathrm{PL}$, with different sign depending on the displacement of the dimer rods (Figure~\ref{Fig3}a). 

To investigate the mechanism leading to the sign flip of $g_{\rm PL}$ for the SLR, we analyze the near-field distribution of the electric field when illuminated with RCP and LCP light. Note that the Lorentz reciprocity theorem relates the outcoupling efficiency to the near-field intensity \cite{novotny2012principles,zhang2023dual}. Figure~\ref{Fig4}b and \ref{Fig4}c show the near-field distribution of the SLR on the xy plane at a height of 10 nm over the top of the rods.  For $|k_x|=0.9 ~\mathrm{rad}/\mu\mathrm{m}$ and a photon energy of 2.11 eV (Figure~\ref{Fig4}b), the field intensity with RCP illumination is stronger than that with LCP illumination. The field intensity with RCP illumination becomes weaker than that with LCP illumination when  $|k_x|=3.1 ~\mathrm{rad}/\mu\mathrm{m}$ and photon energy of 1.98 eV (Figure~\ref{Fig4}c). These results verify that the same SLR mode could possess different handedness under different incident photon energies (and corresponding in-plane wavevectors). In comparison, the field distribution of the quasi-BIC mode is similar for different $k_x$ and photon energies, indicating that this mode preserves its  multipolar character and symmetries along the entire band.

To further support the multipolar-based arguments, we evaluate the helicity density ($h$) maps based on the eigenmodes corresponding to the quasi-BIC and SLR (See Methods and Figure~S7), showing both in-plane and out-of-plane (plane of incidence) cuts: the $xy$ plane crossing the center of the Si rods, and the $xz$ plane crossing the center of the unit cell. A uniform helicity,  mostly concentrated inside, is observed along the rods for the quasi-BIC, with antisymmetric features, being very weak outside the rods. On the other hand, the SLR exhibits a more complex helicity pattern even inside the rods. A clear antisymmetric pattern along the $z$ direction is observed inside each rod, compatible with a strong circulation of the electric field in a $xz$-plane originating an effective $\rm{m}_x$ contribution. Note also that larger helicity densities are found in between the rods. 

\begin{figure}
    \centering
    \includegraphics[width=0.9\linewidth]{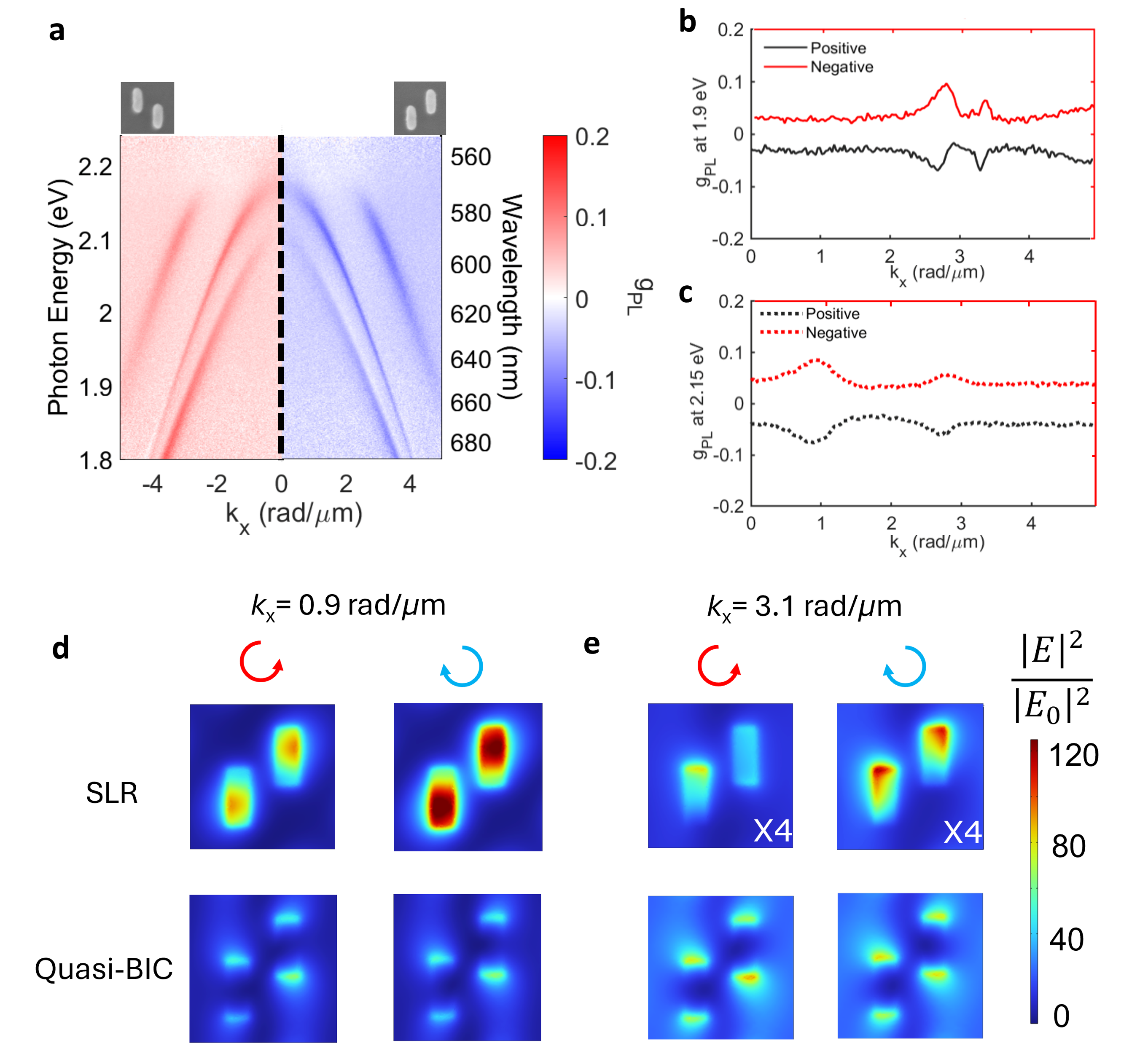}
    \caption{PL dissymmetry factors of the chiral metasurfaces spin-coated with a 380 nm thick dye-doped polymer film. (a) $g_\mathrm{PL}$ dispersion maps as a function of the photon energy and $k_x$ for the metasurface with positively displaced rods (right) and the negatively displaced rods (left). (b) $g_\mathrm{PL}$ as a function of $k_x$ at 1.9 eV, and (c) 2.15 eV.  Electric near-field distribution of the SLR mode and the quasi-BIC mode on the $xy$ plane at a height of 10 nm over the top of the rods with a 380 nm dye-doped polymer film with an illumination at (d) $k_x$ = 0.9$~\mathrm{rad}/\mu\mathrm{m}$ (2.11 eV) and (e) 3.1$~\mathrm{rad}/\mu\mathrm{m}$ (1.98 eV).}
    \label{Fig5}
\end{figure}

We have also explored the impact of the dye layer thicknesses in the $g_\mathrm{PL}$. For this investigation, we prepared a sample with a 380 nm thick dye-doped layer on the same arrays. The corresponding PLE dispersion maps for $k_x$ and $k_y$ are shown in Figure~S8 in the supplementary information, showing a weaker PLE (~$\approx3-5$) in the quasi-BIC mode than the one measured with a thinner dye layer (PLE~$\approx5-8$) (see Figure~S3 in the supplementary information). Furthermore, a waveguide mode arises at higher energies in the $k_x$-dispersion measurements. 
The resulting  $g_\mathrm{PL}$ maps are shown in Figure~\ref{Fig5}a for both positively and negatively displaced nanorods. Apart from  a non-zero background signal, similar quasi-BIC and SLR bands are observed, yielding significant  $g_\mathrm{PL}$ values, along with the new guided mode at higher frequencies beyond the folded diffraction line. The corresponding numerical simulations are shown in Figure~S9  in the supplemental information, which agree well with the behavior of the experimental PL dissymmetry factors. Furthermore, the analysis of the mode decomposition (Figure~S10 in the supplemental information) reveals that the dipole characters of the quasi-BIC mode and the SLR mode are preserved for the thicker layer.  Incidentally, 
the waveguide mode is associated with a significant symmetric $m_x$ component. The helicity densities agree also with the multipolar analysis on the nature of both bands (Figure~S7). The main difference is in the SLR helicity, which is weaker for the thicker layer, revealing in turn a weaker perpendicular confinement and uniform sign. In comparison, the quasi-BIC mode remains essentially unchanged when varying the layer thickness. The most relevant feature is that no sign flip of $g_\mathrm{PL}$ in the SLR mode is observed  in Figure~\ref{Fig5}a for the case of the layer 380 nm thick, unlike that seen for 220 nm thick case (at $|k_x|>2~ \mathrm{rad}/\mu\mathrm{m}$  in Figure~\ref{Fig3}a). According to the near-field distributions (Figure~5d-5e), similar relative intensities are obtained for LCP /RCP illumination at different energies throughout the SLR band, confirming that the sign of  $g_\mathrm{PL}$  remains constant. A possible explanation for the robust values of $g_{\rm PL}$ with layer thickness for the quasi-BIC mode compared to the SLR could be the stronger lateral field confinement,\cite{ter2023direct} also evidenced in our helicity density calculations shown in Figure~S7. The quasi-BIC mode is only sensitive to changes very close to the nanoparticles, so the change in layer thickness does not modify its nature. On the other hand, the SLR mode, which is closer to the diffraction line and less confined, is more sensitive to distance to the interface, thus showing changes in the $g_{\rm PL}$ sign for thin enough layers. Another possible explanation for the robustness of the quasi-BIC mode compared to the SLR lies in their local/nonlocal characteristics \cite{liang2024local}. This is supported by the larger full width at half maximum of the $g_{\rm PL}$ of the quasi-BIC compared with that of the SLR, which indicates greater dissipation due to scattering and outcoupling through localized resonances (Figures~\ref{Fig3}a and \ref{Fig5}a). Owing to the more localized nature of the quasi-BIC mode, its chirality is less sensitive to extrinsic effects and to the layer thickness.

\textit{Conclusions.} The mechanism of CPL generation from quasi-BICs and SLRs in instrinsic chiral Si metasurfaces has been investigated. We have characterized the photoluminescence enhancement and $g_\mathrm{PL}$ in metasurfaces with organic dye layers of different thicknesses. The quasi-BIC mode exhibits high PLE $\approx4-5$ for a layer thickness of 220 nm, but relatively low values of $g_\mathrm{PL}\approx0.1$, which are robust under the variation of $k_x$ and dye thickness. For the SLR mode, higher $g_\mathrm{PL}\approx0.17$ with lower PLE $\approx1.2$ are achieved. In addition, the sign of the $g_\mathrm{PL}$ is inverted under different $k_x$ and dye thickness. Our theoretical analysis based on numerical calculations reveals that the quasi-BIC mode is dominated by strong in-plane antisymmetric electric dipole ($\mathbf{p}$) contributions, leading to significantly enhanced PLE, whereas the SLR mode features pronounced magnetic dipole ($\mathbf{m}$) components that contribute to a higher $g_\mathrm{PL}$, with much lower PLE though. Furthermore, the quasi-BIC mode maintains an invariant spatial distribution of the electric field and helicity density under environmental changes, underscoring its exceptional robustness. In contrast, the SLR mode exhibits significant variation in field distributions under similar changes, leading to the handedness flipping of CPL. These findings elucidate the distinct chiro-optical characteristics of quasi-BIC and SLR modes in intrinsic chiral silicon metasurfaces, with the quasi-BIC mode standing out for its chiro-optical robustness, and offering strategic insights for optimizing CPL performance in metasurface-based luminescent devices.

\section{Methods}

\subsection{Fabrication of Si metasurfaces}
Polycrystalline Si thin films, 90 nm in thickness, were deposited on a synthetic silica glass substrate using low-pressure chemical vapor deposition with SiH$_4$ gas as the Si source. A positive resist (ZEP520A)  was applied to the Si film and subjected to electron-beam lithography. Subsequently, nanorod hole arrays of resist were formed on the Si film through development. Next, a Cr layer (70 nm, by electron-beam deposition) was deposited on the hole array and a lift-off process resulted in the Cr nanorod arrays on the Si film. Using the Cr as a mask, the Si film was then vertically etched by a selective dry etching process (Bosch process) using SF$_6$ and C$_4$F$_8$
 gases. Finally, the Cr mask was removed through wet etching in an acidic solution (S-clean S24, Sasaki Chemical Co., Ltd.). The resulting array covered an area of 2 × 2 mm$^2$.
\subsection{Fourier Microscope} 
Figure~S2 in the supplementary information shows a schematic illustration of the Fourier microscope used to image the angular distribution of PL, which is measured as a function of the photon energy and in-plane wavevector $k$. $k$ is related to the outcoupling angle of the emission from the sample $\theta$ by $k=\frac{\omega}{c}\sin(\theta)$. An optical diffuser is placed in front of the sample to create an unpolarized excitation from the incident beam. An illumination objective is used to excite the sample within the area of the particle array. When the PL from the sample passes through a collection objective, each wave vector component is separated and focused at a different position in the back focal plane (BFP), creating a Fourier image. To measure the CPL, a quarter-wave plate (QWP) and a linear polarizer are placed after the collection objective. Two lenses are used to image the BFP onto the detector. These two lenses are positioned at a focal length after the QWP, increasing the working distance.

\subsection{COMSOL Simulations}
The multipole decomposition and calculation of the near-fields of the modes was performed with the Electromagnetic Waves in Frequency Domain module of COMSOL Multiphysics. The simulation was done for the Si array covered with perylene dye molecules in PMMA. We calculated the polarization $\mathbf{P}$ induced in the system by external plane waves impinging at different angles with TE polarization. The multipole decomposition of the modes can be calculated by integrating $\mathbf{P}$: \cite{Evlyukhin2013,CorbatonVSH,Wu2020c,de2024interaction}


\begin{equation}
    \mathbf{p} = \int \mathbf{P} j_0(kr) d^3\textbf{r}      + \frac{k^2}{2}\int \left \{3 ( \textbf{r} \cdot \textbf{P}) \mathbf{r} - r^2 \mathbf{P} \right) \} \frac{j_2 (kr)}{(kr)^2} d^3\textbf{r}\;,
\end{equation}
\begin{equation}
    \mathbf{m} = - \frac{i 3\omega}{2}\int (\mathbf{r}\times\mathbf{P}) \frac{j_1 (kr)}{kr}d\mathbf{r}\;.
\end{equation}
$\omega$ is the optical frequency. Higher-order contributions are negligible in the considered spectral range.

The helicity density is defined as:

\begin{equation}
    h(\mathbf{r}) = -\frac{1}{2}\frac{\sqrt{\varepsilon_0\mu_0}}{\omega}\mathrm{Im}\{\mathbf{E}^*\cdot\mathbf{B} \},
\end{equation}
where $\varepsilon_0$ and $\mu_0$ are the vacuum permittivity and vacuum permeability, respectively. The vectors $\mathbf{E}$ and $\mathbf{B}$ denote the electric and magnetic fields, respectively.

\subsection{Simulations of Angle-Dependent $g_\mathrm{PL}$ Maps}
Based on the Lorentz reciprocity theorem that the light out-coupling efficiencies associated with the near-field enhancements,~\cite{zhang2023dual,ramezani2016modified} we calculated the near-field electric field intensity in the organic layer. To obtain the $g_\mathrm{PL}$, we use LCP and RCP as the light source to illuminate the metasurface. The molecular dipole orientations are assumed to be fully random. We obtain the $g_\mathrm{PL}$ as a function of $k$ and $\lambda$ by evaluating the field intensity difference between LCP and RCP excitations as follows,\cite{maksimov2014circularly,cen2024chirally}.
\begin{equation}
    g_\mathrm{PL}(k,\lambda)=~2\iiint_V \frac{|\mathbf{E}_\mathrm{LCP}(\mathbf{r},k,\lambda)|^2-|\mathbf{E}_\mathrm{RCP}(\mathbf{r},k,\lambda)|^2}{|\mathbf{E}_\mathrm{LCP}(\mathbf{r},k,\lambda)|^2+|\mathbf{E}_\mathrm{RCP}(\mathbf{r},k,\lambda)|^2}d\mathbf{r}^3,
\end{equation}
where $k$ and $\lambda$ are the in-plane wave vector and the emission wavelength, respectively. $\mathbf{E}_\mathrm{LCP}(\mathbf{r},k,\lambda)$ and $\mathbf{E}_\mathrm{RCP}(\mathbf{r},k,\lambda)$ are the near-electric field at position $\mathbf{r}$, for wave vector $k$, and wavelength $\lambda$ with LCP and RCP excitation, respectively. The integrated volume $V$ covers the whole organic layer.

\begin{acknowledgement}
This project was funded by the European Union. Views and opinions expressed are however those of the author(s) only and do not necessarily reflect those of the European Union or the European Innovation Council and SMEs Executive Agency (EISMEA). Neither the European Union nor the granting authority can be held responsible for them. (SCOLED, Grant Agreement No. 101098813) Y.-C.W. acknowledges support from the National Science and Technology Council (NSTC) through the postdoctoral research abroad program (113-2917-I-564-036). J.L.P., A.G.M. and J.A.S.G. acknowledge financial support from the grants TED2021-131417B-I00 (BICPLAN6G), TED2021-130786B-I00, PID2020-113533RB-C33, PID2021-126046OB-C22, and PID2022-137569NB-C41 (LIGHTCOMPAS), funded by MCIN/AEI/10.13039/501100011033, “ERDF A way of making Europe”, and European Union NextGenerationEU/PRTR. J.L.P. also acknowledges the financial support from a Margarita Salas contract CONVREC-2021-23 (University of Valladolid and European Union NextGenerationEU), and the Advanced Materials programme supported by MCIN with funding from European Union NextGenerationEU (PRTR-C17.I1). B.C.L.d.L. acknowledges a predoctoral contract funded as well by “ESF Investing in your future”  (PREP2022-000426).
S.M. acknowledges financial support from JSPS, Japan (25K01501, 23K23044, 22K18884, JPJSBP120239921).
\end{acknowledgement}

\begin{suppinfo}

The Supporting Information is available free of charge.
\end{suppinfo}


\bibliography{achemso-demo}

\end{document}